\documentclass[aps,prb,twocolumn,showpacs,groupedaddress,floatfix]{revtex4}

\usepackage{amsmath}
\usepackage{amsfonts}
\usepackage{amssymb}
\usepackage{graphicx}

\begin{document}

\title{Testing the Berry phase model for extraordinary Hall effect in SrRuO$_{3}$}
\date{\today }

\begin{abstract}
Recently it has been suggested that the complicated temperature dependence of
the extraordinary Hall effect (EHE) in the itinerant ferromagnet SrRuO$_{3}$
could be explained by the Berry phase effect in the crystal momentum space. We test
this model by measurements of EHE as a function of an applied magnetic field at a
constant temperature and show that the results seem to contradict the Berry
phase mechanism.
\end{abstract}


\author{Yevgeny Kats}
\altaffiliation{Present address: Department of Physics, Harvard University, Cambridge, MA 02138.}
\author{Isaschar Genish}
\author{Lior Klein}
\affiliation{Department of Physics, Bar-Ilan University, Ramat-Gan 52900, Israel}
\author{James W. Reiner}
\altaffiliation{Present address: Department of Applied Physics, Yale University, New Haven, CT 06520-8284.}
\author{M. R. Beasley}
\affiliation{T. H. Geballe Laboratory for Advanced Materials, Stanford University, Stanford, California 94305, USA}
\maketitle

The Hall effect in magnetic materials includes, in addition to an
\textit{ordinary} (or \textit{regular}) Hall effect (OHE), which
originates from the Lorentz force and depends on the magnetic induction
$\mathbf{B}$, an \textit{ extraordinary} (or \textit{anomalous})
Hall effect (EHE), which depends on the magnetization $\mathbf{M}$.
Usually, the EHE is attributed to spin-dependent scattering, and the
total Hall effect is given by
\begin{equation}
\rho_{xy} = \rho_{xy}^{OHE} + \rho_{xy}^{EHE} = %
R_{0}B_{z} + R_{s}(\rho)\mu_{0}M_{z}\text{,} \label{Hall effect}
\end{equation}
where $R_{0}$ is the ordinary Hall coefficient related to the carrier density
$n$, and $R_{s}$ is the extraordinary Hall coefficient, which depends on the
resistivity $\rho$ as $R_{s} = a\rho + b\rho^{2}$, where the linear term is
due to a spin-dependent preferred direction in scattering (``skew scattering"),%
\cite{Smit} and the quadratic term is due to a lateral displacement involved in
the scattering (``side jump").\cite{Berger}

Recently, it has been suggested that the Berry phase effect\cite{Berry} in the
crystal momentum space ($\mathbf{k}$ space) can also give rise to EHE.%
\cite{Jungwirth, Fang, Yao} This is an intrinsic effect, which does not
involve scattering, but it depends on the Bloch states and their occupation.
In this model, the EHE is described as
\begin{equation}
\rho_{xy}^{EHE}=-\rho^{2} \sigma_{xy}^{BP}(M)\text{,}
\label{EHE-Berry}
\end{equation}
where the Berry phase transverse conductivity $\sigma_{xy}^{BP}(M)$ does not
depend on $\rho$, and the dependence of $\sigma_{xy}^{BP}$ on $M$ should be
calculated from the band structure. First, this mechanism was invoked to
explain the EHE in (III,Mn)V ferromagnetic semiconductors,\cite{Jungwirth} then
in SrRuO$_{3}$ (Ref. \onlinecite{Fang}) (which is the subject of the current paper), and
later it was shown that the Berry phase effect in $\mathbf{k}$ space can be the
dominant mechanism even in iron.\cite{Yao} Actually such mechanism for the EHE
had been suggested by Karplus and Luttinger\cite{Karplus-Luttinger} a long time
ago, but it was disregarded later. This effect should be distinguished,
however, from the Berry phase effect related to a motion in a topologically
nontrivial spin background in \textit{real} space, which has been also proposed
as a source of EHE for some materials.\cite{real-space-Berry-EHE}

\begin{figure}[ptb]
\includegraphics[scale=0.51, trim=140 320 180 -150]{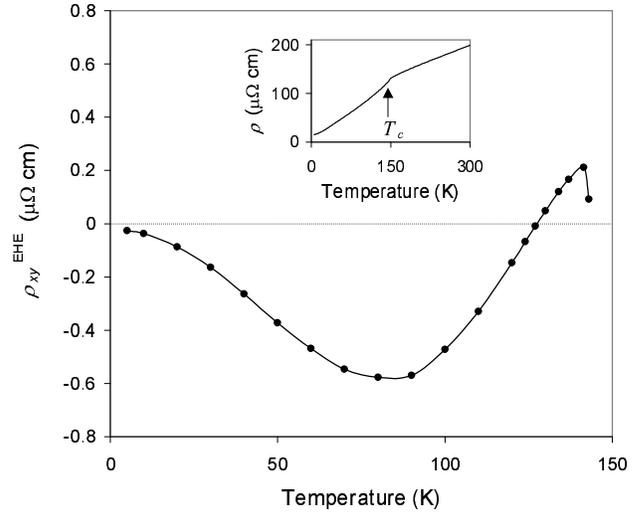}
\caption{Extraordinary Hall effect $\rho_{xy}^{EHE}$ (due to the spontaneous
magnetization) as a function of temperature. The inset shows the
longitudinal resistivity $\rho$ as a function of temperature.}
\label{EHE vs T}
\end{figure}

The EHE in the $4d$ itinerant ferromagnet SrRuO$_{3}$ exhibits
a nonmonotonic temperature dependence, including a change of
sign\cite{Izumi, Klein-EHE} (see Fig. \ref{EHE vs T}), and $R_s$ does
not follow the relation $R_s = a\rho + b\rho^{2}$.\cite{Klein-EHE}
Fang \textit{et al.}\cite{Fang} argued that this behavior can be explained
by the Berry phase effect in $\mathbf{k}$ space, which predicts a peculiar
nonmonotonic dependence of $\sigma_{xy}^{BP}$ on $M$. The authors supported
their contention by band calculations, which predicted EHE of a correct order
of magnitude and roughly reproduced its temperature dependence.
However, band calculations for SrRuO$_{3}$ are very sensitive to the input
parameters.\cite{Fang, band-SRO} Therefore, while the calculations support the
explanation, they leave open the possibility that in practice the Berry phase
effect in SrRuO$_{3}$ is much smaller, and the EHE is caused by a different
mechanism.

Another point which raises questions regarding the applicability of the
calculations of Fang \textit{et al.} is their assumption that the exchange
band splitting vanishes at $T_{c}$. In many itinerant ferromagnets, and
probably also in SrRuO$_{3}$,\cite{Dodge} the band splitting does not
disappear at $T_{c}$, but at a temperature higher by an order of
magnitude. In such materials, at $T_{c}$, the magnetization disappears on the long
scale, but the short-range order remains, and the spin-split bands are quite well
defined locally (see, e.g., Ref. \onlinecite{local-band}). According to the
calculation of Fang \textit{et al.}, the EHE changes sign when the band
splitting is about one-third of its zero-temperature value, which probably does
not happen below $T_{c}$.

The experiment presented here explores the changes in EHE resulting from
changes in $M$ due to a magnetic field applied at a fixed temperature.
This allows us to test the applicability of the Berry phase model directly,
by a comparison between temperature-dependent and field-dependent behavior,
without making assumptions regarding the details of the band structure.
Particularly, we inquire whether the quantity that vanishes at
$T \simeq 127$ K in Fig. \ref{EHE vs T} is the $\rho$-dependent $R_{s}$ from
Eq. (\ref{Hall effect}) or the $M$-dependent $\sigma_{xy}^{BP}$ from
Eq. (\ref{EHE-Berry}).

We study epitaxial films of SrRuO$_{3}$ grown by reactive electron beam
coevaporation\cite{films growth} on miscut ($\sim2^{\circ}$) SrTiO$_{3}$
substrates. The films are single phase, with a single easy axis of
magnetization roughly at $45^{\circ}$ out of the plane of the film (the
direction of the easy axis varies slightly as a function of temperature).%
\cite{Klein-Mag} The film whose results are presented here has a thickness of
$30$ nm, and $T_{c} \simeq 147$ K. The films were patterned by
photolithography. The current path was perpendicular to the easy axis.
The residual longitudinal offset in Hall effect measurements
was canceled by repeating the measurements with a reversed magnetic field and
taking half the difference of the results. All measurements were performed with
the films uniformly magnetized, including at zero magnetic field.

\begin{figure}[ptb]
\includegraphics[scale=0.46, trim=170 330 200 -170]{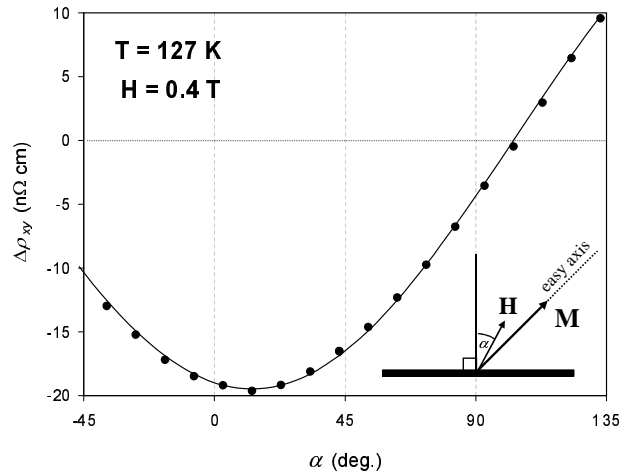}
\caption{Field-induced contribution to the Hall effect at $T = 127$ K, $H =
0.4$ T, as a function of the direction of the field (the angle $\alpha$ is
shown at the right bottom part of the figure). The solid curve is a fit from
which the OHE contribution was evaluated.}%
\label{OHE-EHE}
\end{figure}

In order to separate the OHE contribution, we measured the Hall effect at a
low magnetic field ($H \le 0.4$ T) as a function of the direction of the
field. In such fields, the change in $\rho_{xy}$ is linear in $H$, implying
that the change in $\rho_{xy}^{EHE}$, if it is significant, is also linear
in $H$. In addition, for such fields $\mathbf{M}$ does not rotate away from
the easy axis because the anisotropy field is of order of $10$ T.%
\footnote{For example, if a field of $0.4$ T is applied at $60^{\circ}$
relative to the easy axis, the magnetization will rotate by about $1^{\circ}$,
so that the error due to incorrect modeling of the \textit{field-induced}
change in the Hall effect is only about $4\%$ of it. The additional error due
to change in the \textit{zero-field} EHE can be made smaller than that error if
the measurement is performed at a temperature where the zero-field EHE is
small, as we did ($T = 127$ K).} %
Since the easy axis is at $\alpha \simeq 45^{\circ}$ ($\alpha$ is defined at
the right bottom part of Fig. \ref{OHE-EHE}), the EHE and the OHE contributions
have different symmetries, and can be separated. Particularly, the EHE
contribution should not be affected at all when the magnetic field is applied
perpendicularly to the easy axis. In general, we expect
\begin{equation}
\Delta\rho_{xy} = R_0 H \cos\alpha + \frac{d\rho_{xy}^{EHE}}{dM}\chi
H\cos(\alpha-\alpha_{ea}), \label{OHE-EHE-fit}
\end{equation}
where $\alpha_{ea}$ is the direction of the easy axis, $\chi$ is the
susceptibility, and $\rho_{xy}^{EHE}(\rho,M)$ is considered to be a function
of $M$ alone, since for constant temperature and magnetization direction $\rho$
is a function of $M$ (Lorentz magnetoresistance is negligible at the temperature
of our measurement).\cite{Kats-MR} Figure \ref{OHE-EHE} shows the additional
Hall effect (i.e., after subtracting the EHE measured at zero field) as a
function of the direction of the field, and a fit according to Eq.
(\ref{OHE-EHE-fit}).
It turns out that the parts of the OHE and the EHE in the field-induced Hall
effect are comparable in magnitude. For a field applied along the easy axis:
$(60 \pm 3)\%$ of the change in Hall effect was due to the OHE, while $(40 \pm
3)\%$ was due to the EHE.

\begin{figure}[ptb]
\includegraphics[scale=0.48, trim=140 320 180 -170]{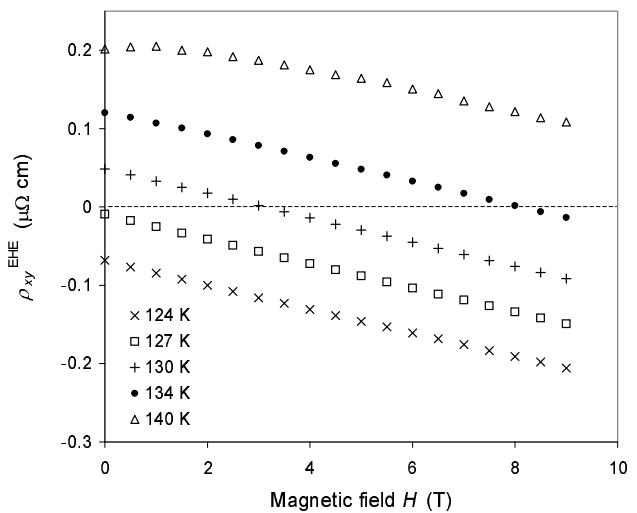}
\caption{Extraordinary Hall effect as a function of the magnetic
field $H$ at several temperatures (indicated in the legend).}
\label{EHE vs H}
\end{figure}

Figure \ref{EHE vs H} shows the EHE as a function of the magnetic field $H$ at
different temperatures, after subtracting the OHE contribution. (The field
was applied along the easy axis, in order to create maximal possible changes
in $M$.)
Interestingly, while the magnetization increases with the increasing field,
not only does the EHE decrease, it even changes sign. Furthermore, EHE exists
even at $T = 127$ K, where the zero-field $R_{s}$ (see Fig. \ref{EHE vs T})
vanishes.

These results seem to qualitatively agree with the predictions of the Berry
phase model for these temperatures, since by applying a magnetic field we reach
values of $M$ that at zero field exist at lower temperatures:
Figure \ref{EHE vs T} implies that in the range of temperatures presented in
Fig. \ref{EHE vs H}, $|\sigma_{xy}^{BP}(M)|$ decreases with increasing $M$;
therefore, the EHE is expected to decrease when a magnetic field is applied.

\begin{figure}[ptb]
\includegraphics[scale=0.5, trim=140 320 180 -170]{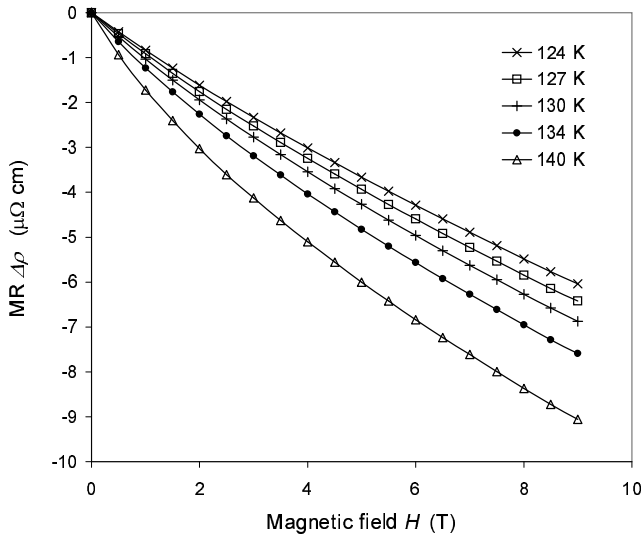}
\caption{Magnetoresistance $\Delta\rho(H)=\rho(H)-\rho(0)$ as a function of
the magnetic field $H$, corresponding to the measurements presented in
Fig. \ref{EHE vs H}.}
\label{MR vs H}
\end{figure}

On the other hand, the increase in $M$ diminishes magnetic scattering,
resulting in a negative magnetoresistance (MR) $\Delta\rho(H) = \rho(H)-\rho(0)$
(see Fig. \ref{MR vs H}).

\begin{figure}[ptb]
\includegraphics[scale=0.46, trim=140 340 180 -160]{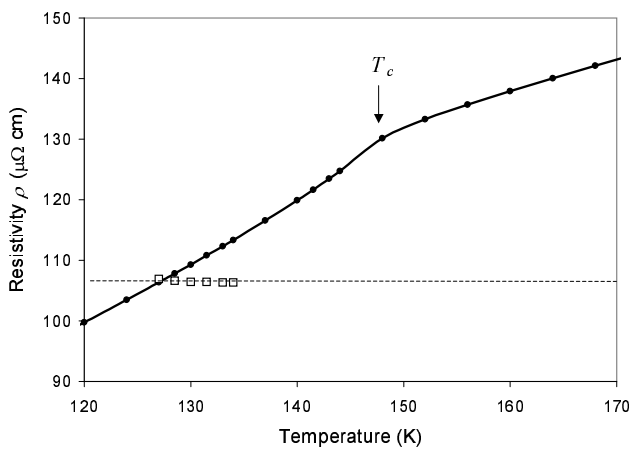}
\caption{The solid curve shows the temperature dependence of the zero-field
resistivity $\rho$, and the squares denote the resistivity for which EHE
vanishes at applied magnetic field, as a function of the temperature at which
the field is applied.}
\label{r0 vs T}
\end{figure}

Thus the results can qualitatively agree also with the prediction based on
Eq. (\ref{Hall effect}), since by applying a magnetic field we attain lower
resistivities $\rho$, and in our range of temperatures, $R_{s}$ decreases with
decreasing resistivity (see Fig. \ref{EHE vs T} and inset).

Quantitative examination of the results supports the second possibility. It
turns out, for example, that the MR ($\simeq -7$ $\mu\Omega$ cm) which is
required at $T = 134$ K to make the EHE vanish brings the resistivity to the
zero-field resistivity of $T = 127$ K (where the EHE vanishes at zero field).
Figure \ref{r0 vs T} shows this pattern for a range of temperatures:
the EHE always vanishes at the same value of $\rho$.
This behavior is consistent with Eq. (\ref{Hall effect}).

The vanishing of EHE at constant resistivity cannot be consistent with Eq.
(\ref{EHE-Berry}), since the identical resistivities do not correspond
to identical values of $M$. While direct magnetic measurements could be
advantageous, accurate magnetization measurements of thin films are
plagued by big substrate contributions. On the other hand, it is possible
to analyze the question by considering changes in magnetic scattering
involved in our experiment.

From a temperature $T > 127$ K, vanishing EHE can be achieved either by
lowering the temperature to $127$ K or by applying an appropriate magnetic
field. In both cases $\rho$ decreases to the same value. However, in the first
case, the decrease in $\rho$ is partly related to a decrease in non-magnetic
scattering (phonons, etc.), while in the second case the whole change
in $\rho$ is due to change in magnetic scattering.\cite{Kats-MR} Therefore,
the magnetic scattering is different in the two cases, indicating different
values of $M$. Thus, it is not a particular value of $M$ in $\sigma_{xy}^{BP}(M)$,
which makes EHE vanish.

Quantitatively, we estimate that the non-magnetic part of $d\rho/dT$
around $130$ K is about $0.50$ $\mu\Omega$ cm/K, which is the value of
$d\rho/dT$ above $T_{c}$ where the magnetic resistivity
saturates. Therefore, non-magnetic resistivity plays an important
role. For example, only $3.5$ $\mu\Omega$ cm of the $7$ $\mu\Omega$ cm
difference in the zero-field resistivity between $134$ and $127$ K is due to
magnetic resistivity. The magnetic resistivity of $127$ K is achieved at
$134$ K already for $H = 3.4$ T (this is the field for which the MR is
$3.5$ $\mu\Omega$ cm, see Fig. \ref{MR vs H}), while the EHE vanishes
only at $H = 8.1$ T.

In conclusion, when temperature-dependent and field-dependent measurements of
the EHE are combined, the results cannot be simply explained in terms of the
Berry phase model. On the other hand, it seems that Eq. (\ref{Hall effect})
describes the EHE correctly, although the microscopic origin of the
$\rho$-dependence of $R_{s}$ remains unclear.

We appreciate valuable comments from J. S. Dodge. We acknowledge support by
the Israel Science Foundation founded by the Israel Academy of Sciences and
Humanities.


\begin{thebibliography}{99}

\bibitem{Smit} J. Smit, Physica (Amsterdam) \textbf{24}, 39 (1958).

\bibitem{Berger} L. Berger, Phys. Rev. B \textbf{2}, 4559 (1970).

\bibitem{Berry} M. V. Berry, Proc. R. Soc. Lond. \textbf{A392}, 45 (1984).

\bibitem{Jungwirth} T. Jungwirth, Q. Niu, and A. H. MacDonald, Phys. Rev. Lett. \textbf{88}, 207208 (2002).

\bibitem{Fang} Z. Fang, N. Nagaosa, K. S. Takahashi, A. Asamitsu, R. Mathieu, T. Ogasawara, H. Yamada, M. Kawasaki, Y. Tokura, and K. Terakura, Science \textbf{302}, 92 (2003).

\bibitem{Yao} Y. Yao, L. Kleinman, A. H. MacDonald, J. Sinova, T. Jungwirth, D.-S. Wang, E. Wang, and Q. Niu, Phys. Rev. Lett. \textbf{92}, 037204 (2004).

\bibitem{Karplus-Luttinger} R. Karplus and J. M. Luttinger, Phys. Rev. \textbf{95}, 1154
(1954); J. M. Luttinger, \textit{ibid.} \textbf{112}, 739 (1958).

\bibitem{real-space-Berry-EHE}
 J. Ye, Y. B. Kim, A. J. Millis, B. I. Shraiman, P. Majumdar, and Z. Tesanovic, Phys. Rev. Lett. \textbf{83}, 3737 (1999);
 Y. Taguchi, Y. Oohara, H. Yoshizawa, N. Nagaosa, and Y. Tokura, Science \textbf{291}, 2573 (2001);
 R. Shindou and N. Nagaosa, Phys. Rev. Lett. \textbf{87}, 116801 (2001).

\bibitem{Izumi} M. Izumi, K. Nakazawa, Y. Bando, Y. Yoneda, and H. Terauchi, J. Phys. Soc. Jpn. \textbf{66}, 3893 (1997).

\bibitem{Klein-EHE} L. Klein, J. W. Reiner, T. H. Geballe, M. R. Beasley, and A. Kapitulnik, Phys. Rev. B \textbf{61}, R7842 (2000).

\bibitem{Kats-MR} Y. Kats, L. Klein, J. W. Reiner, T. H. Geballe, M. R. Beasley, and A. Kapitulnik, Phys. Rev. B \textbf{63}, 054435 (2001).

\bibitem{band-SRO} G. Santi and T. Jarlborg, J. Phys.: Condens. Matter \textbf{9}, 9563 (1997).

\bibitem{Dodge} J. S. Dodge, E. Kulatov, L. Klein, C. H. Ahn, J. W. Reiner, L. Mieville, T. H. Geballe, M. R. Beasley, A. Kapitulnik, H. Ohta, Yu. Uspenskii, and S. Halilov, Phys. Rev. B \textbf{60}, R6987 (1999).

\bibitem{local-band} V. Korenman, J. L. Murray, and R. E. Prange, Phys. Rev. B \textbf{16}, 4032 (1977);
                     V. Korenman and R. E. Prange, Phys. Rev. Lett. \textbf{53}, 186 (1984).

\bibitem{films growth} S. J. Benerofe, C. H. Ahn, M. M. Wang, K. E. Kihlstrom, K. B. Do, S. B. Arnason, M. M. Fejer, T. H. Geballe, M. R. Beasley, and R. H. Hammond, J. Vac. Sci. Technol. B \textbf{12}, 1217 (1994).

\bibitem{Klein-Mag} L. Klein, J. S. Dodge, C. H. Ahn, J. W. Reiner, L. Mieville, T. H. Geballe, M. R. Beasley, and A. Kapitulnik, J. Phys.: Condens. Matter \textbf{8}, 10111 (1996).

\end{thebibliography}
\end{document}